\documentclass[runningheads]{svmult}

\usepackage{makeidx}   % allows index generation
\usepackage{graphicx,rotate}  % standard LaTeX graphics tool
\usepackage{subeqnar}  % subnumbers individual equations
\usepackage{multicol}  % used for the two-column index
\usepackage{physprbb}  % modified textarea for proceedings,
\makeindex             % used for the subject index
\newcommand{\ergacf}{{\rm\thinspace erg}}

\newcommand{\Kacf}{{\rm\thinspace K}}
\newcommand{\keVacf}{{\rm\thinspace keV}}
\newcommand{\kmacf}{{\rm\thinspace km}}
\newcommand{\kpcacf}{{\rm\thinspace kpc}}

\newcommand{\Msunacf}{\hbox{$\rm\thinspace M_{\odot}$}}

\newcommand{\sacf}{{\rm\thinspace s}}
\newcommand{\yracf}{{\rm\thinspace yr}}

\newcommand{\ergpsacf}{\hbox{$\ergacf\sacf^{-1}\,$}}
\newcommand{\kmpsacf}{\hbox{$\kmacf\sacf^{-1}\,$}}
\newcommand{\Msunpyracf}{\hbox{$\Msunacf\yracf^{-1}\,$}}

\begin{document}

\title*{Cooling flows in clusters of galaxies}

\author{A.C. Fabian}

\institute{Institute of Astronomy, Madingley Road, Cambridge CB3 0HA, UK}

\maketitle              % typesets the title of the contribution

\begin{abstract}
The gas temperature in the cores of many clusters of galaxies drops
inward by about a factor of three or more within the central 100~kpc
radius. The radiative cooling time drops over the same region from 5
or more Gyr down to about $10^8\yracf$. Although it would seem that
cooling has taken place, XMM and Chandra spectra show no evidence for
strong mass cooling rates of gas below 1--2 keV. Chandra images show
holes coincident with radio lobes and cold fronts indicating that the
core regions are complex. The observational situation is reviewed here
and ways in which continued cooling may be hidden are discussed,
together with the implications for any heat source which balances
radiative cooling.
\end{abstract}

\section{Introduction}

The gas density within the central 100 kpc or so of the centre of most
clusters of galaxies is high enough that the radiative cooling time of
the gas is less than $10^{10}$~yr. The cooling time drops further at
smaller radii, suggesting that in the absence of any balancing heat
source much of the gas in the central regions is cooling out of the
hot intracluster medium. In order to maintain the pressure required to
support the weight of the overlying gas, a slow, subsonic inflow known
as a cooling flow develops.

X-ray observations made before Chandra and XMM-Newton were broadly
consistent with the cooling flow picture (see 23 for a review and 42
for an opposing view), although several issues remained unresolved.
The first issue was the observed X-ray surface brightness profile,
which was not as peaked as expected from a homogeneous flow. Instead a
multiphase gas was assumed, dropping cold gas over a range of radii.
The second was the fate of the cooled gas. At the rates of 100s to
more than $1000\Msunpyracf$ found in some clusters, the central
galaxies should be very bright and blue if the cooled gas forms stars
with a normal intial-mass-function. In many cases they do have excess
blue light indicative of massive star formation [36, 1, 13, 15], but
at rates which are a factor of 10 to 100 times lower than the X-ray
deduced mass cooling rate. It has been argued [46] that there is no
significant sink in terms of cold gas clouds. A third issue involved
the shape of the soft X-ray spectrum, which was inconsistent with a
simple cooling flow. Absorption intrinsic to the flow was found to be
a possible explanation [2, 3]. A final, major, issue was whether the
neglect of heating is justified. The effect of gravitational heating
as the gas flows was taken into account, but the effects pf any
central radio source, which pumps energy into the surrounding gas via
jets, together with disturbances due to subclusters plunging into the
core every few Gyr were not included due to a lack of quantitative
information. Heat flow due to thermal conduction was also generally
assumed negligible.

The situation with cluster cooling flows has been clarified over the
past year, particularly by the high spatial resolution imaging of
Chandra and the high spectral resolution of the XMM-Newton Reflection
Grating Spectrometer (RGS). Chandra images show much detail in the
cores of clusters, with bubbles from radio sources [41, 24] and cold
fronts [40, 57] seen. RGS spectra [50, 54, 38] confirm the presence
of a range of temperatures in cooling flow clusters but fail to show
evidence for gas cooling below 1--2~keV. Simply put, the data are
consistent with gas cooling at a high rate to about one third of the
mean temperature beyond 100~kpc but then vanishing.

At about the same time, the evidence for both warm [35,17, 20] and
cold [19] molecular gas at the centres of cooling flows clusters has
become widespread. In some extreme cases there may be over
$10^{11}\Msunacf$ of cold gas [19]. The presence of dust in these
regions is also widespread, as demonstrated by the Balmer decrement in
the optical/UV nebulosities commonly seen (e.g. 32, 15), dustlanes,
and submm and IR detections [18, 3, 34].  It is therefore possible
that more star formation, and in particular cold gas clouds, may be
found in and around central cluster galaxies (see also 29 for a
discussion of the properties of very cold gas clouds). There has also
been the intriguing detection of OVI emission from A2597 with FUSE
[47]. Lastly, recent numerical simulations of evolving
cluster which include radiative cooling of the gas predict cooling
flows (e.g. 48).

At face value the X-ray data tempt many to assume that some form of
heating balances cooling and so dismiss cooling flows altogther. That
ignores the how, why and what of the heating, which remains unsolved,
although several candidates have been identified [56, 6, 16, 10, 11].
Some form of feedback is probably required to prevent all of the gas
from being heated up. If feedback does occur we have a good chance to
observe how it works, since the region is spatially resolved and
optically thin. The process is of wide importance, since it provides
the upper mass limit for galaxies (in simulations of the galaxy
luminosity function, [39] switch off cooling in massive galaxies).

My own view is to treat it as an intriguing astrophysical puzzle whch
can be tackled observationally.  Heating from radio sources and
infalling subclusters must occur at some level, but whether it can
balance radiative cooling over the required spatial scales to better
than a factor of a few is not yet clear. Cooling probably does account
for the observed star formation and cold gas clouds. A major remaining
issue is whether the mass cooling rates are reduced from the earlier
X-ray deduced rates by a factor of a few, ten or a hundred. We may be
witnessing a nearby example of the kind of feedback processes common
in galaxy formation; in particular, one in which accretion onto the
central black hole and the resultant kinetic energy release play a
major role.

\section{Chandra results}

Chandra images show structure in cluster cores. The X-ray emission is
steeply peaked into the centres of many clusters but there are holes
and fronts in the peak. Markevitch et al [40] found sharply defined
cold fronts on A2142, across which the pressure is continuous yet the
temperature changes by a factor of about 2. Ettori \& Fabian [21]
note that thermal conduction must be heavily suppressed in order that
such sharp features can last long enough to be common. The fronts probably
indicate that the gas of subclusters does not readily mix with the
existing intracluster medium, presumably because they are separate
magnetic structures. Also the cores of infalling subclusters may not
be strongly shocked in decelerating into the core (see [44, 22]).

\begin{figure}
\begin{center}
\includegraphics[width=.7\textwidth]{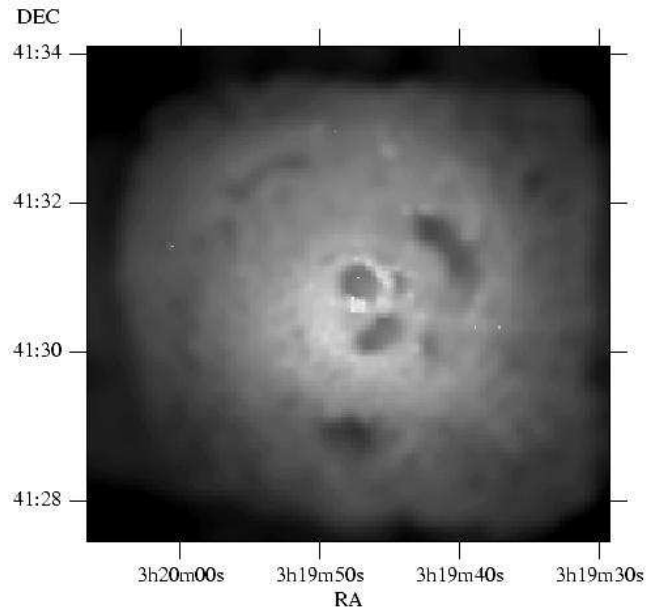}
\includegraphics[width=.7\textwidth]{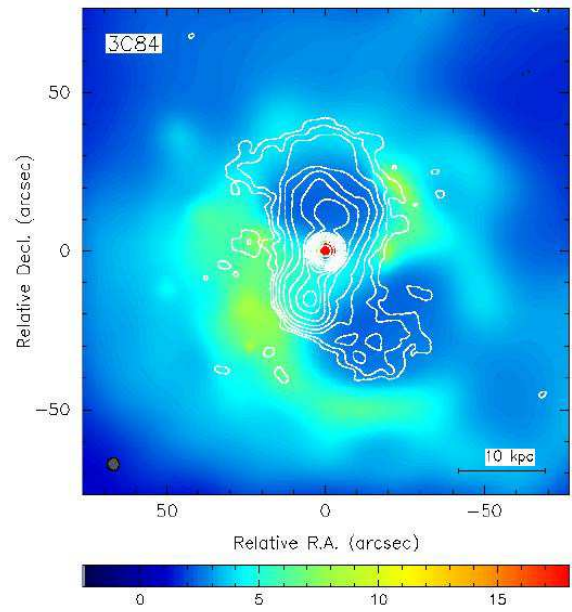}
\end{center}
\caption{(Top) Adaptively-smoothed 0.5--7~keV ACIS-S X-ray image
of the centre of the Perseus cluster. (Bottom) Radio image (1.4~GHz
restored with a 5 arcsec beam, produced by G. Taylor; see [24],
overlaid on an adaptively smoothed 0.5--7~keV X-ray map.}
\end{figure}

\begin{figure}
\begin{center}
\includegraphics[width=.6\textwidth]{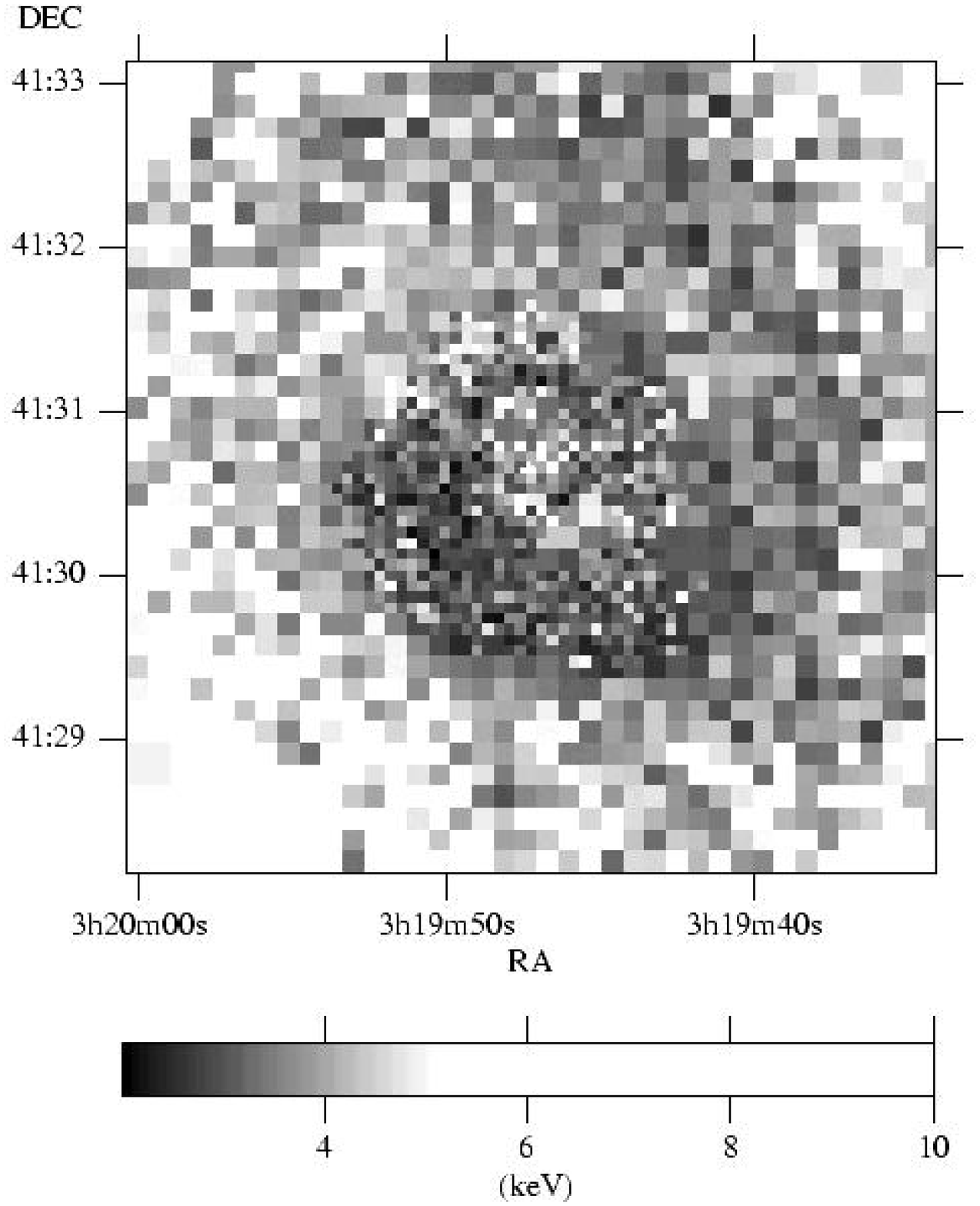}
\includegraphics[width=.6\textwidth]{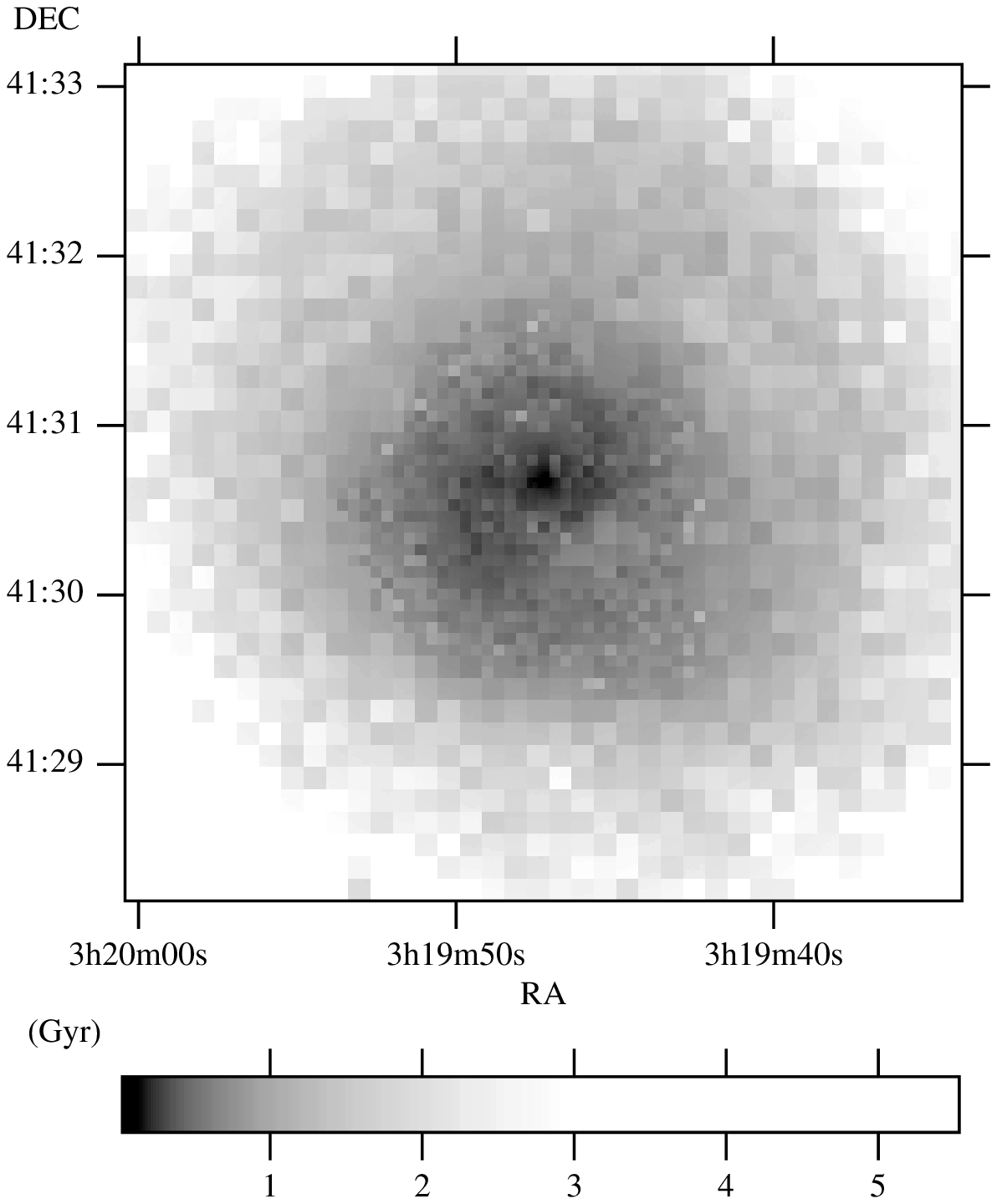}
\end{center}
\caption{(Top) Temperature and (Bottom) radiative cooling time maps
produced from the X-ray colour ratios [24]. Note that
the coolest gas ($T\sim 2.5\keVacf$) with the shortest cooling time
($\sim 0.3$~Gyr) lies in the rim around the N lobe and in the E bright
blob. Single-phase gas has been assumed for the analysis. }
\end{figure}

\subsection{The Perseus cluster}

Holes in the X-ray surface brightness are seen to coincide with some
radio lobes. The best examples are in the Perseus cluster, and were
first seen with ROSAT [7]. Chandra shows that
they have bright rims of X-ray {\it cool} gas [7].
This is contrary to the work of [33] who predicted that
the rims would signify shocks. Other holes coincident with radio lobes
are found in Hydra A [41, 16] and many
other clusters. The puzzling aspect if radio sources are heating the
cooling gas is that in all cases reported the {\it coolest} gas seen
is that closest to the radio lobes. Of course there is much energy
going into the lobes, but the energy from the $PdV$ work expended in
forming the holes can propagate away as sound waves, and the
relativistic energy stored in the bubbles can be lifted away
and out of the immediate core by buoyancy.

Provided that the filling factor of the holes by relativistic plasma
is high, then the jet power required to make the holes in the Perseus
cluster is considerable at about $10^{45}\ergpsacf$ [27].

\subsection{A1795, A2199 and the Centaurus cluster}

\begin{figure}
\begin{center}
\includegraphics[width=.7\textwidth]{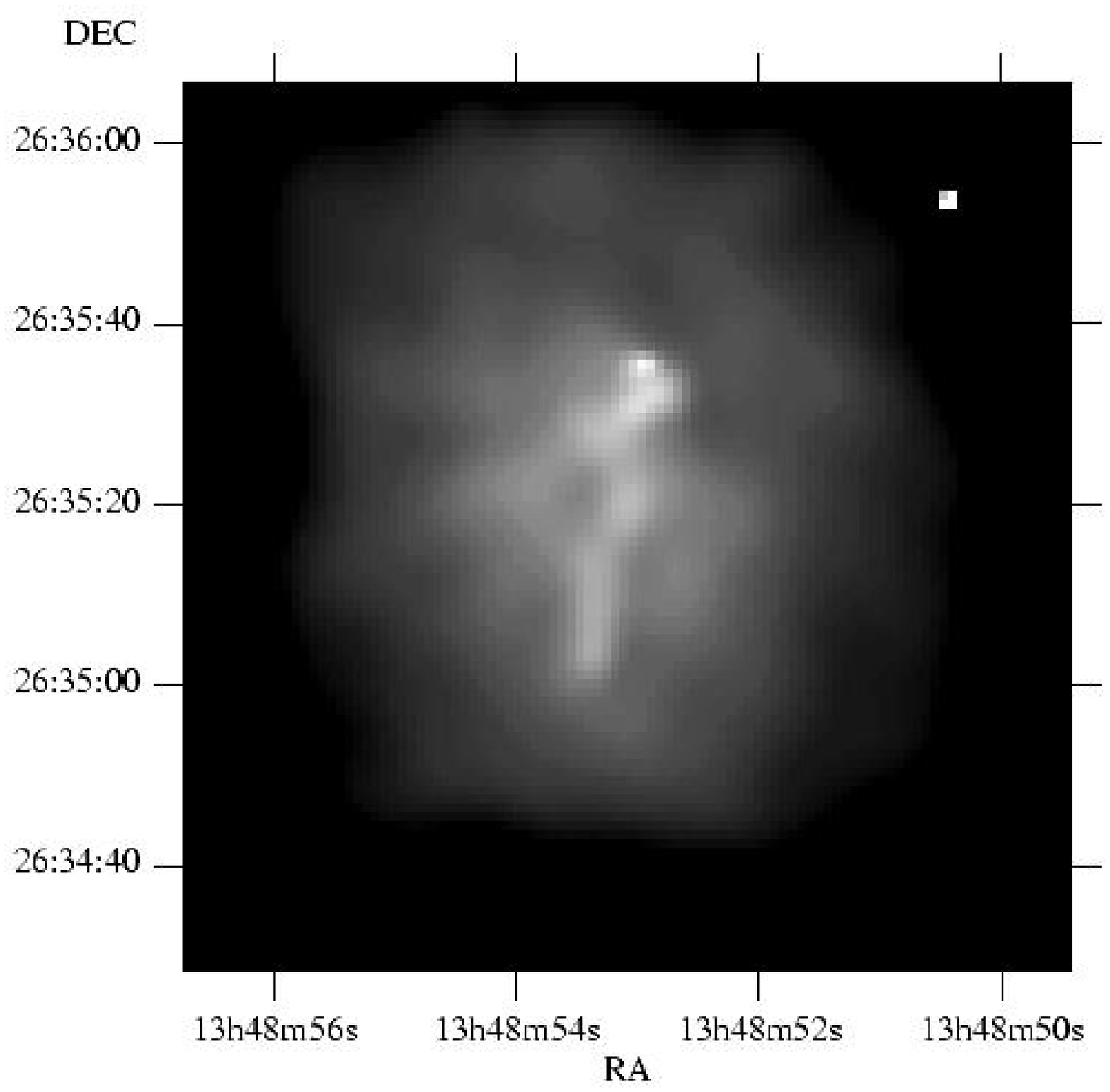}
\includegraphics[width=.7\textwidth]{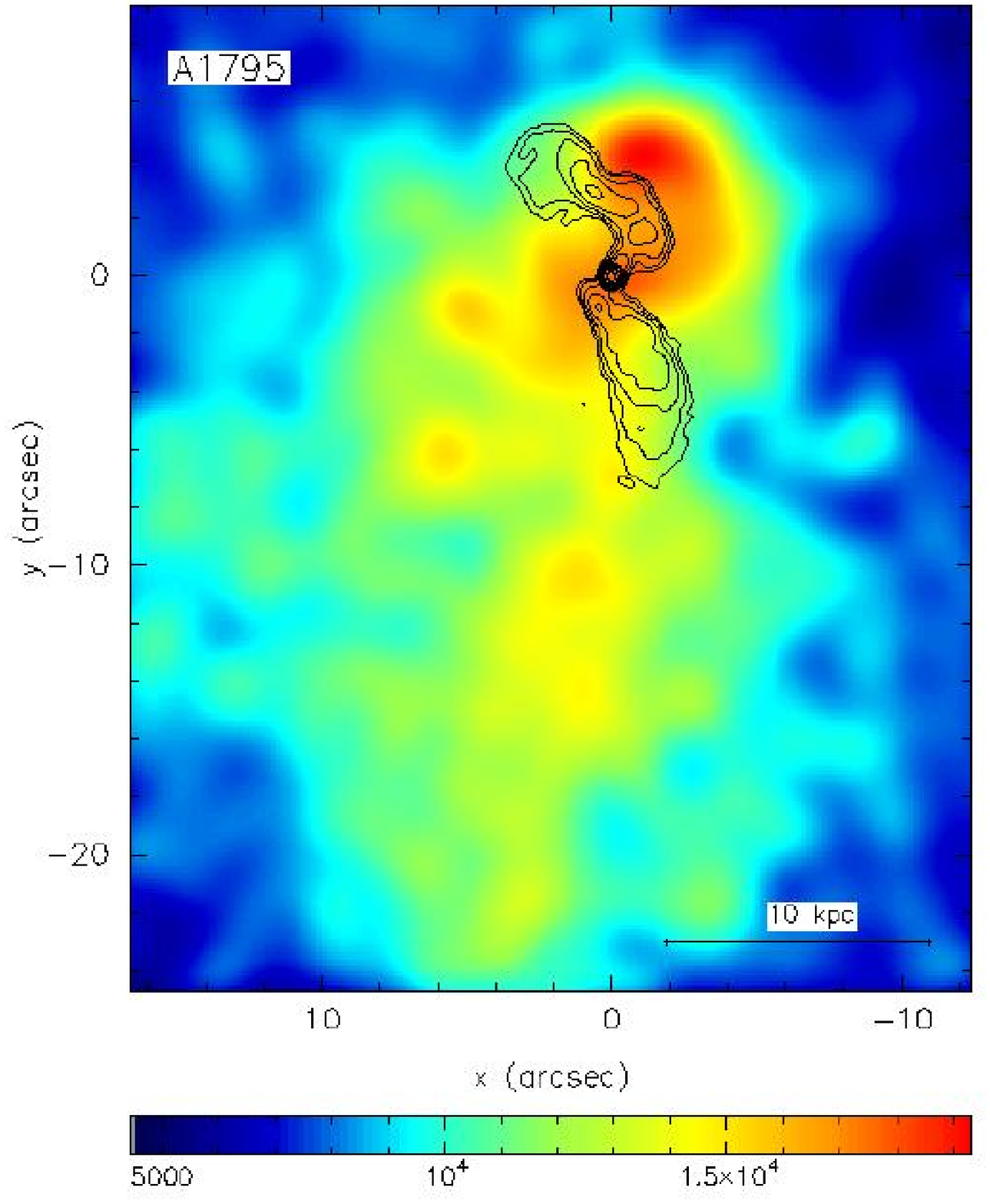}
\end{center}
\caption{(Top) Adaptively-smoothed X-ray image of the centre of A1795
[25]. (Bottom) Overlay of the 3.6~cm radio emission [31] on the X-ray image.}
\end{figure}

\begin{figure}
\begin{center}
\includegraphics[width=.7\textwidth]{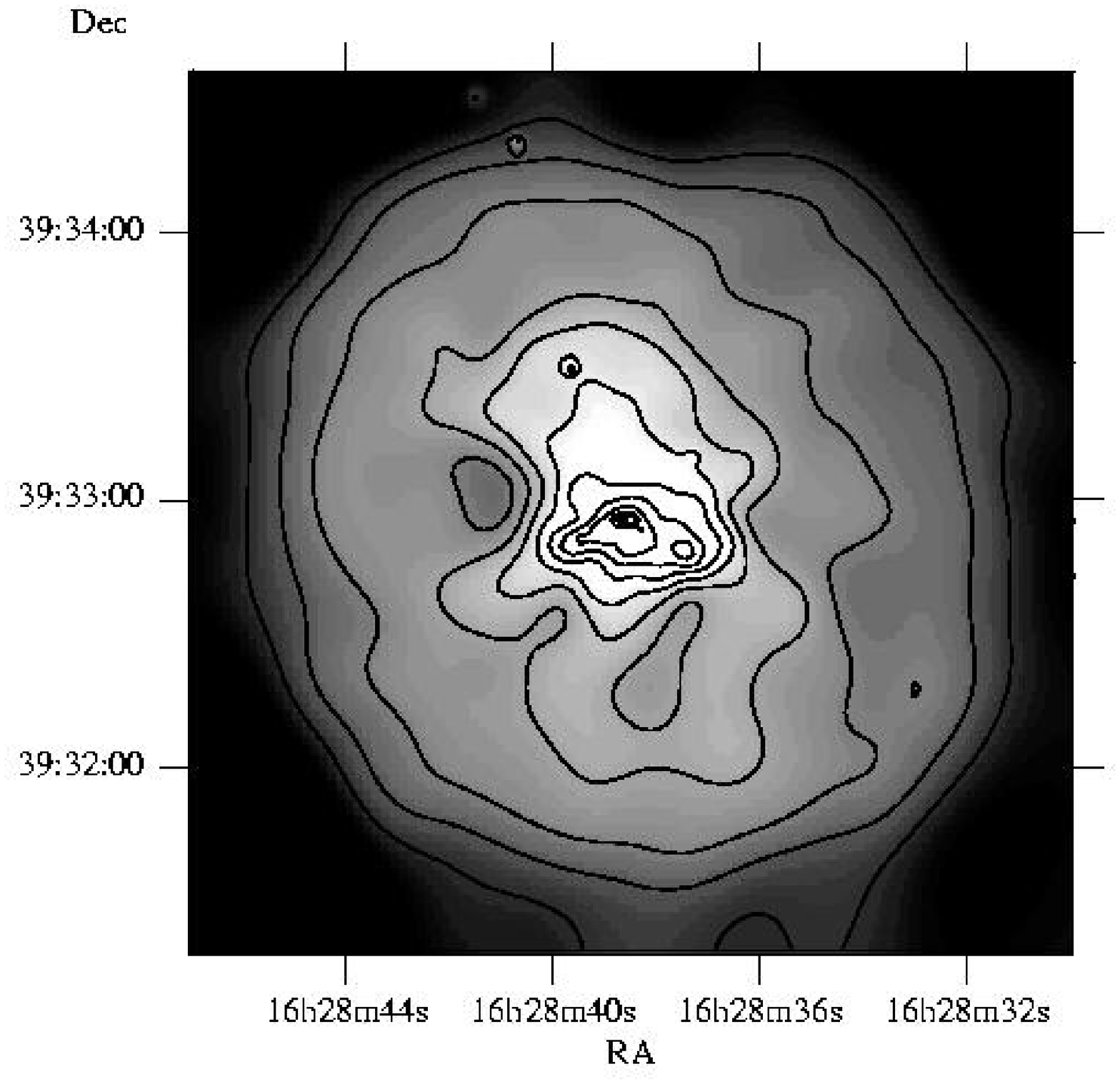}
\includegraphics[width=.7\textwidth]{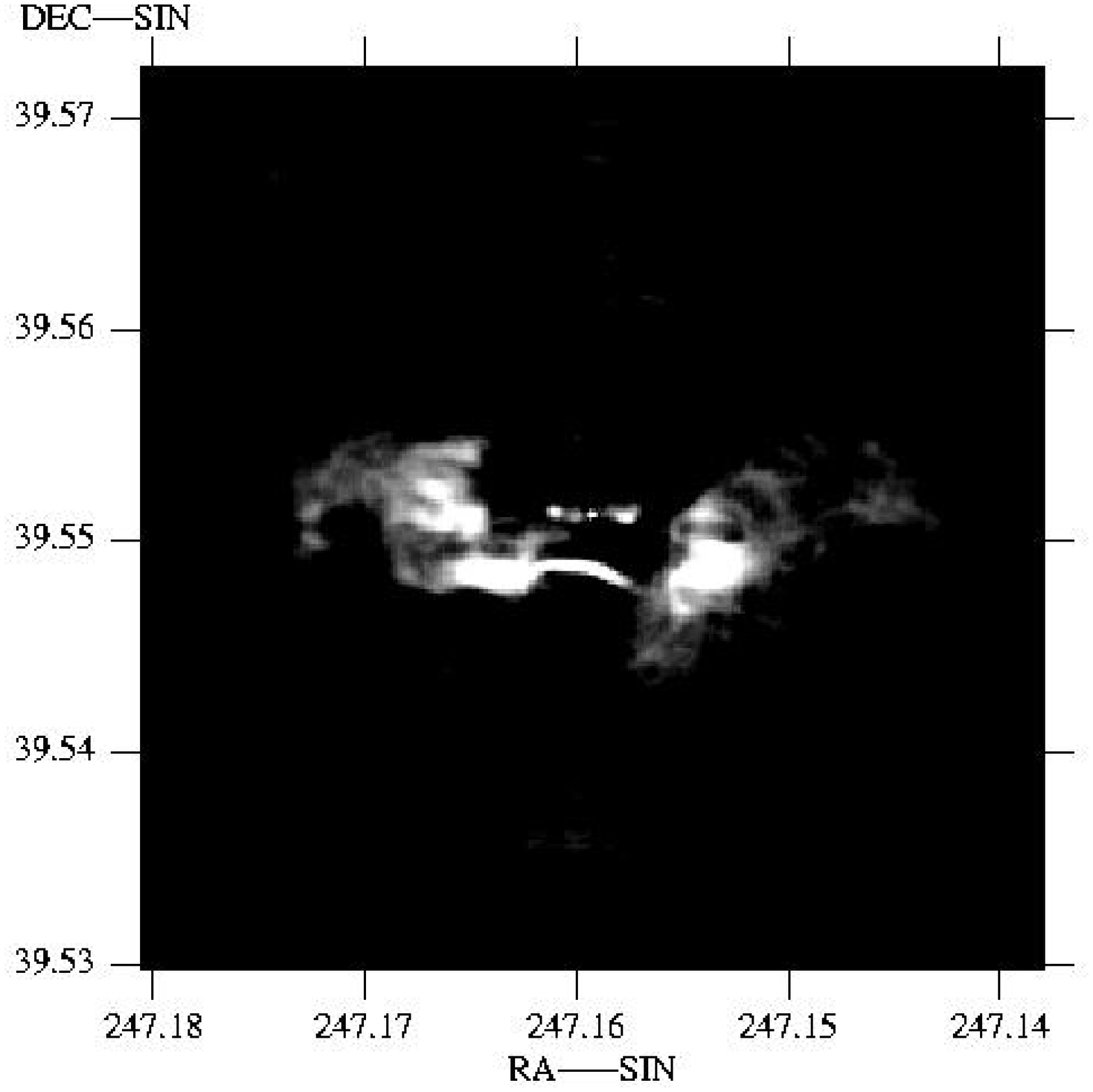}
\end{center}
\caption{(Top) Adaptively-smoothed 0.5--3~keV X-ray image of the core
of A2199 (the contours are logarithmic); (Bottom) 1.7~GHz radio image
[30]. See [37] for analysis of
the X-ray image).the X-ray surface brightness drops at the position of
the outer radio lobes.}
\end{figure}

\begin{figure}
\begin{center}
\includegraphics[width=.6\textwidth]{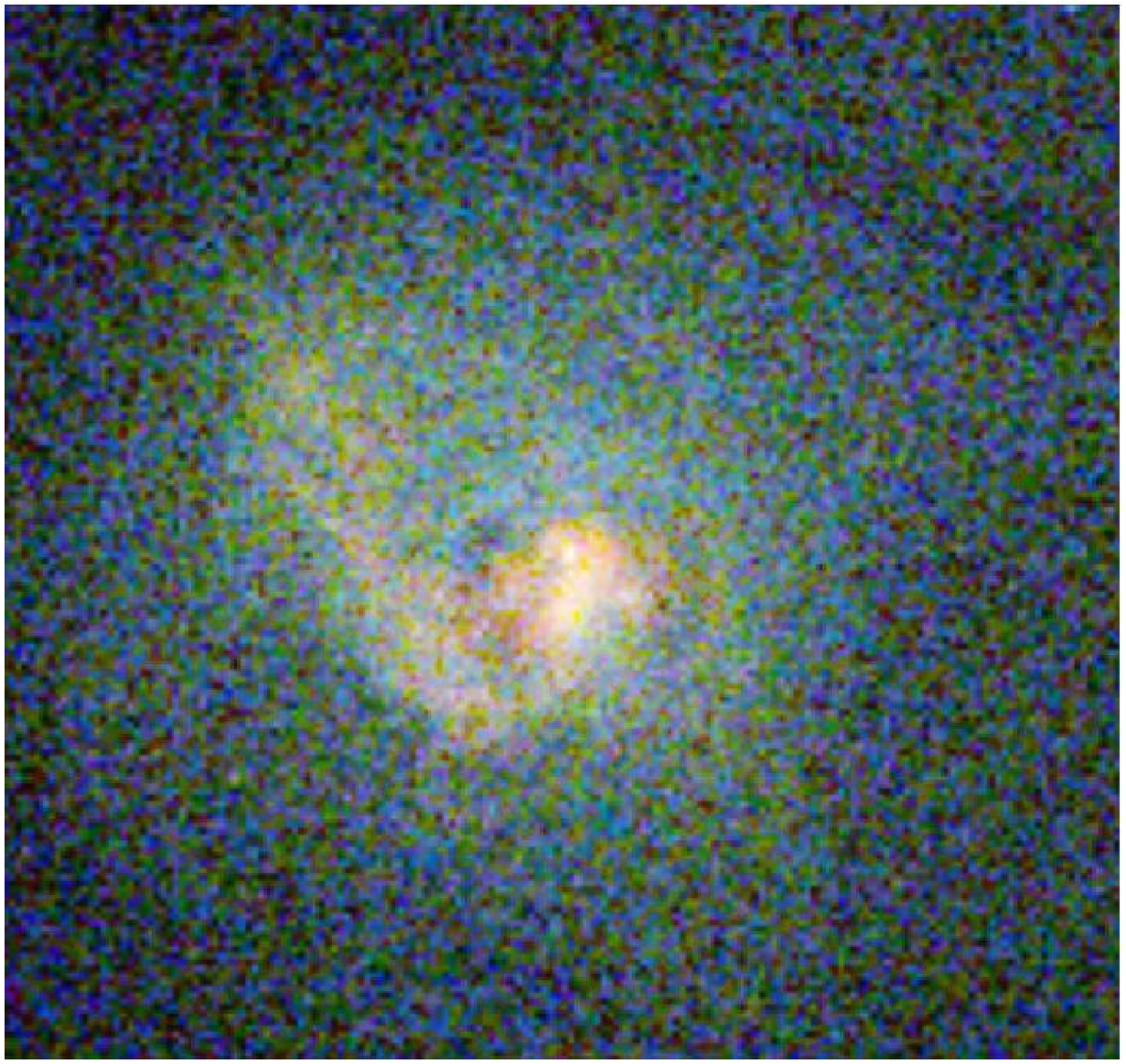}
\includegraphics[width=.7\textwidth]{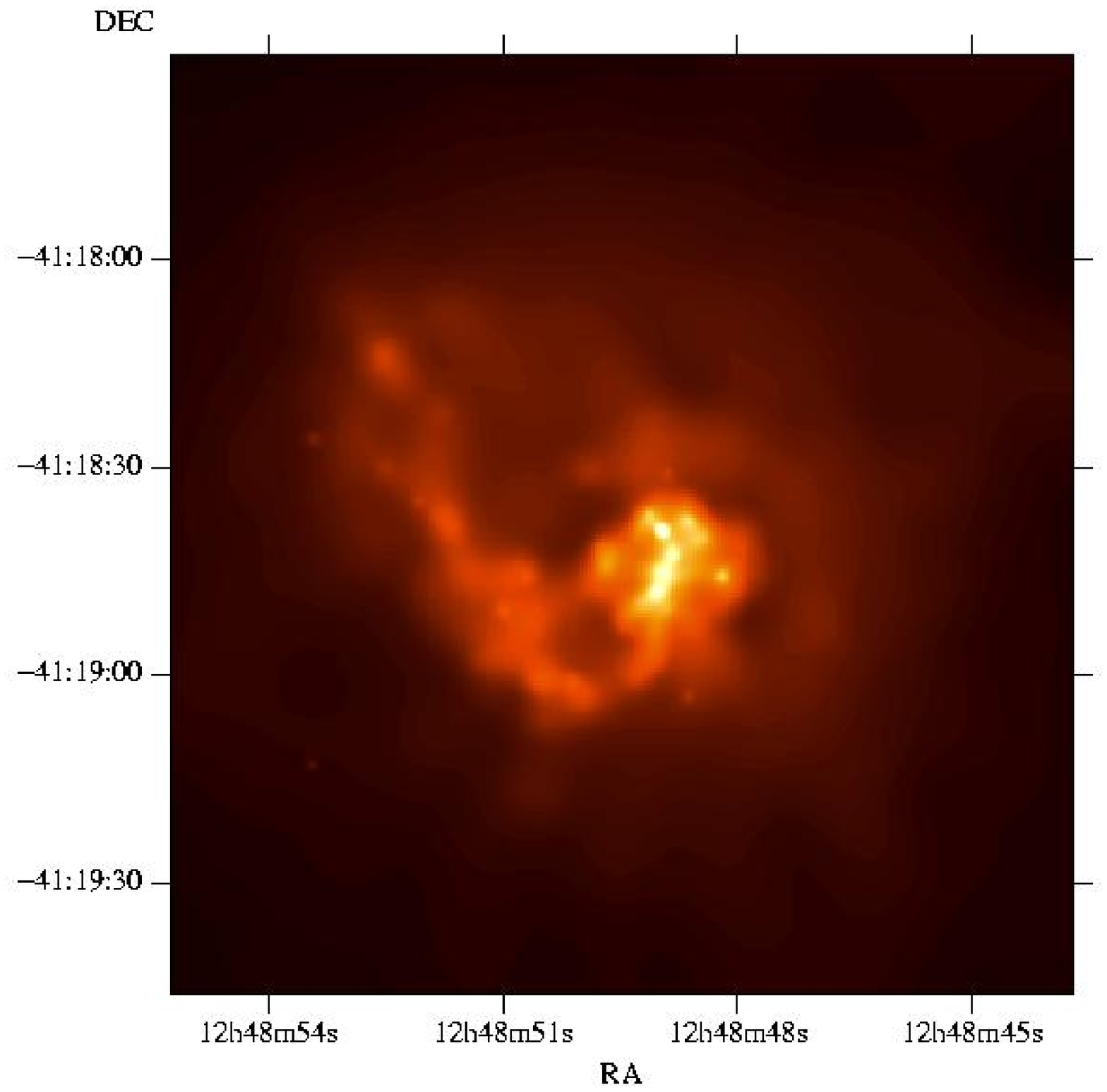}
\end{center}
\caption{(Top) Raw colour image 
and (Bottom) adaptively-smoothed 0.5--3~keV X-ray image of the core
of the Centaurus cluster [52]. The radio source in
the central cluster galaxy, NGC\,4696, has a complex cap-like
structure, 30 arcsec wide, 
fitting around the head of the X-ray plume-like feature [55].}
\end{figure}

The Chandra image of A1795 [25] shows an 80~kpc long
soft X-ray feature coincident with an H$\alpha$ filament found by
[14]. It is plausibly a cooling wake trailing behind the
central galaxy, which is at the head of the filament. The galaxy is
moving around in the core of the cluster at a few hundred $\kmpsacf$.
There is no evidence from the temperatures of the gas that this motion
has heated the gas significantly. The central galaxy in A2199 may also
be oscillating, as deduced from the unusual morphology of its radio
source [12]. Again, the coolest gas appears to be close
to both the radio source and the central galaxy [37].

The Centaurus cluster shows a cool, plume-like feature which appears
about 20 kpc long [52]. The gas temperature drops
from about 3.5~keV down to about 1 keV at the centre, with evidence
that more than one phase is present there. Interestingly, the known
strong abundance gradient in iron peaks at about 1.5 times Solar at
about 15~kpc before dropping back down to about 0.5 near the centre.
This limits the amount of widespread mixing or convection that can
have occurred within the core.

In summary, the high-spatial-resolution Chandra data are revealing
that cluster cores are complex with a combination of holes due to
radio lobes as well as  filaments, plumes and cold fronts. All cores
studied so far where the radiative cooling time of the gas is a few
Gyr show significant central temperature drops. 

\begin{figure}

\begin{center}
\rotatebox{270}{
\includegraphics[width=.8\textwidth]{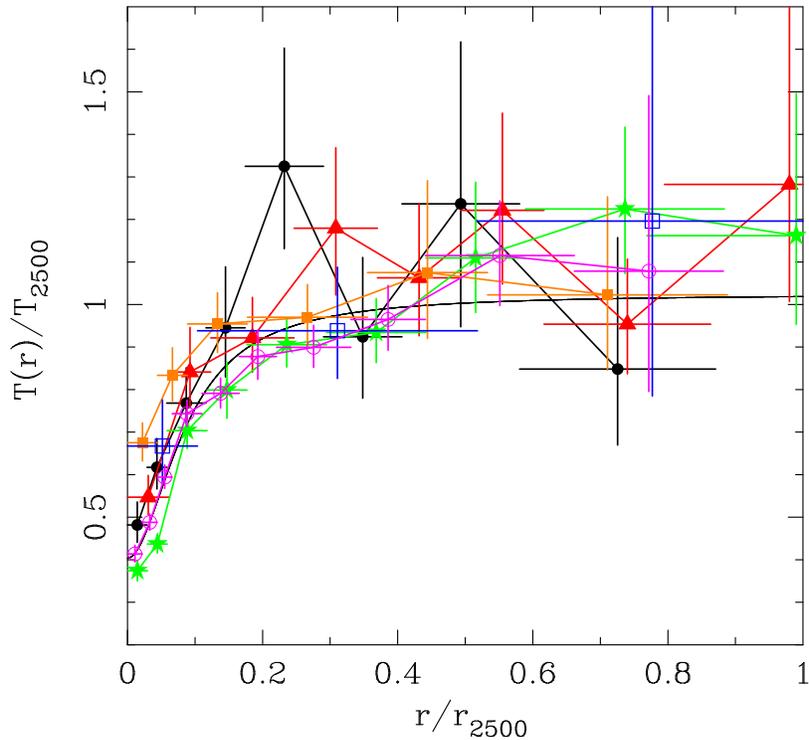}}
\end{center}
\caption{Chandra temperature profiles for six massive clusters: PKS0745-191, 
A2390, A1835, MS2137-2353, RXJ1347-1145 and 3C295, from [4]. Note the
temperature drop within $\sim0.2 r_{2500}\sim 120\kpcacf$.}
\end{figure}

\section{XMM-Newton results}

\begin{figure}
\begin{center}
\includegraphics[width=\textwidth]{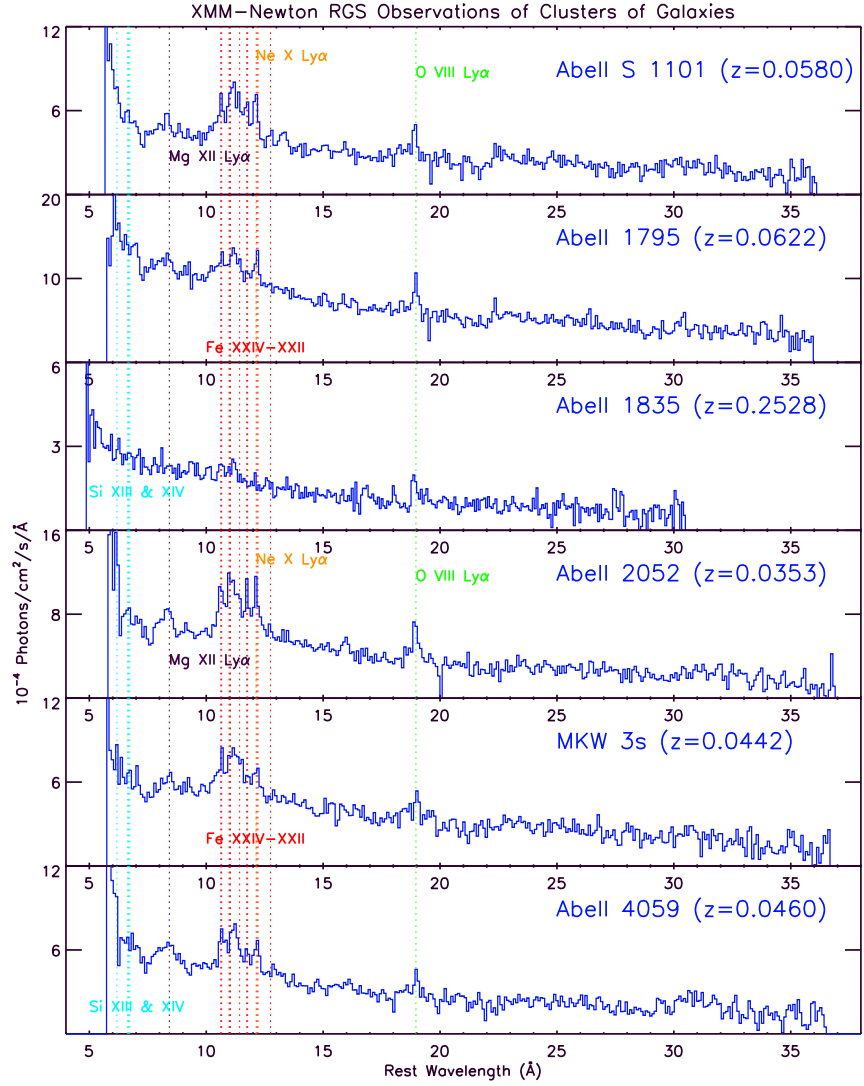}
\end{center}
\caption{RGS spectra of 6 cooling flow clusters, kindly provided by J.
Peterson and J. Kaastra. Emission lines between 10-13~A indicate the
presence of cooler gas in these clusters (at 1--3~keV) but the lack of
lines between 13 and 18~A shows that gas is not radiatively cooling
below 1--2~keV at high rates in any simple unobscured manner. }
\end{figure}

The most striking results have come from the RGS data which show
little evidence for gas cooling below 1--2~keV [50, 53, 38]. Emission
lines from FeXX and XVII, for example, should be bright and easily
seen if the mass cooling rate is high, but they are absent. EPIC CCD
spectra (e.g. [9, 43])
confirm this result.

Various explanations have been given [50, 26, 28]. The gas may be
cooling and yet appear to vanish when it reaches say 2~keV. Clouds of
cold gas may photoelectrically absorb the soft X-rays, or the gas may
have become dense enough to separate from the flow and mix in with
surrounding hotter gas [45]. Alternatively it may
mix in with colder gas (for example that associated with the optical
filaments or the molecular gas -- note that this also explains why the
filaments are so bright; [28]).

Another possibility is that the metals in the gas are not uniformly
mixed in, but have a bimodal distribution [26]. Gas in which ten per
cent has a metallicity of 3 times solar and 90 per cent has zero
metallicity has the same spectrum as gas at 0.3 solar if cooling is
unimportant. When it does cool however, the metal emission lines cool
only 10 per cent of the gas and so are much reduced as compared with
the situation if they were responsible for cooling all the gas.

If heating is the explanation, then it cannot just be some low level
of heat which stops the gas at 1--2~keV, since it would then
accumulate at that temperature, contrary to observation. It has to
halt the cooling over the full range of temperatures, and thus radii.
How this can happen is a puzzle. If the radio source is responsible,
then it may be intermittent. Maybe we do not see the heating phase,
which is short lived. However the power required to stem the flow
during the heating phase then goes up to high values. There may not be
any problem in the Virgo cluster around M87 where the energy
rquirements are relatively small, but in a massive cluster like A1835
[50, 53] the necessary power may exceed $10^{46}\ergpsacf$ [26].

\section{Discussion}

The central 100 kpc radius region in most clusters has a radiative
cooling time shorter than 5 Gyr (Peres et al 1997) and many have
central cooling times of only $10^8\yracf$. The gas temperature
drops by a factor of 3 or more over this radius range. It seems
plausible that the temperature drop and short radiative cooling times
are related and that the low temperatures are caused by radiative
cooling.

It is then a puzzle as to why gas which has cooled by a factor of
three, and for which the radiative cooling time has reduced by a
factor of ten or more, is not seen to cool further.

There are two obvious solutions, both of which have some difficulties.
The first solution is that the gas does cool but either the soft X-ray
emission is absorbed or the cooling is non-radiative and due, say, to
mixing. The problem of the fate of cooled gas then remains. The second
solution is that some heating balances cooling. The problem here is
that the heat has to balance cooling over a wide range of radii and a
wide range of timescales. Also observations of radio lobes which are a
likely source of heat indicate that they coincide with the coolest gas
in cluster cores.

The answer may be more complex, with the major temperature drop being
due to a combination of in situ radiative cooling and gas introduced
from dense cooling subclusters. Heat from the kinetic energy of
infalling subclusters, and turbulence, continues to be dissipated
throughout the core, reducing the age of any steady central cooling
region to only a few Gyr. An intermittent central radio source powered
by accretion from the intracluster medium heats and churns up some of
the coolest gas at the centre. Radiatively-cooling clumps (possibly
metal rich) fall out of the mean slow inflow once their temperature
drops to below one third of the outer temperature and rapidly mix with
cooler gas clouds closer to the centre. The mixture would have a
temperature of about $10^5\Kacf$ (as in mixing layers; Begelman \&
Fabian 1990), and rapidly lose its thermal energy by UV emission. Most
of this would be absorbed by neighbouring cold gas and dust, to be
reradiated as optical/UV line emission and infrared dust emission
[28]. Massive star formation take place in cooled clumps and spreads
dust into the surrounding gas, further enhancing cooling via infrared
emission.

\section{Acknowledgements}

I am grateful to my many collaborators, in particular Steve Allen,
Carolin Crawford, Stefano Ettori, Roderick Johnstone, Jeremy Sanders,
Robert Schmidt and Greg Taylor for help and discussions, and the
organisers for creating a timely and memorable meeting. The Royal
Society is thanked for support.

%INDEX%%%%%%%%%%%%%%%%%%%%%%%%%%%%%%%%%%%%%%%%%%%%%%%%%%%%%%%%%%%%%%%
% Please check with the editor of your book whether he plans to
% include a "mutual" subject index - if so, please code your entries
% in the standard syntax. For your own purposes you may print your
% "personal" index by using the following commands:
%
%\clearpage
%\addcontentsline{toc}{section}{Index}
%\flushbottom
%\printindex
%%%%%%%%%%%%%%%%%%%%%%%%%%%%%%%%%%%%%%%%%%%%%%%%%%%%%%%%%%%%%%%%%%%%%

\end{document}